\documentclass[10pt,preprint]{aastex7}
\newcommand{\roma}[1]{\uppercase\expandafter{\romannumeral#1}}

\usepackage{amsmath}
\usepackage{subfigure}
\usepackage{hyperref}

\usepackage{lineno}
\usepackage{color}
\usepackage{makecell}
\usepackage{graphicx}
\usepackage{natbib}
\usepackage{amssymb,txfonts}
\usepackage{multirow}
\usepackage{array}
\usepackage{sidecap}
\usepackage{hyperref}
\usepackage{epstopdf}
\usepackage{footnote}
\usepackage{tabularx}
\usepackage{booktabs}

\shorttitle{Fast reconnection in a coronal torn plasma sheet}
\shortauthors{Tang et al.}

\begin{document}

\title{Fast reconnection in a coronal torn plasma sheet}
\correspondingauthor{Zehao Tang}

\author{Zehao Tang}

\affiliation{Yunnan Observatories, Chinese Academy of Sciences, Kunming 650216, China}
\affiliation{University of Chinese Academy of Sciences, Beijing, 100049, China}
\affiliation{Yunnan Key Laboratory of Solar Physics and Space Science, Kunming 650216, People's Republic of China}
\email[show]{tangzh@ynao.ac.cn}

\begin{abstract}
	
	Tearing instability, also known as plasmoid instability, is an effective mechanism to speed up magnetic reconnection process, working in a wide range of magnetized plasma systems with different spatial scales, ionization degrees, and collisionality. However, due to observational limitations, observations of {plasma sheet} tearing and the resulting plasmoids are rather scarce. This scarcity significantly hinders our understanding of the role of plasmoids in the reconnection process from an observational perspective. Using high-spatiotemporal multiwavelength observations from the Solar Dynamics Observatory, we traced the entire evolution of a coronal {plasma sheet}. Its formation was driven by the emergence of photospheric magnetic flux, followed by tearing, and eventual decay. The evolution of the {plasma sheet} exhibited two distinct stages. Initially, it rose rapidly, lengthened, and underwent tearing at a low frequency. Subsequently, its ascent slowed, it began to shorten, and the tearing occurred more frequently. Detailed analysis of the reconnecting {plasma sheet} focuses on heating, plasmoid dynamics (formation and ejection), and the resulting reconnection rate change. Two key heating processes are identified: {plasma sheet} tearing and coalescence involving plasmoids and magnetic cusps. More importantly, combining observations with analytical studies suggests that plasmoids are key carriers of magnetic flux fast transferring in the observed torn {plasma sheet}, and their formation and ejection significantly enhance the reconnection rate and facilitate the onset of fast reconnection.

\end{abstract}

\section{Introduction}

A current sheet is defined as a layer where current highly concentrates. Current sheets contain magnetic tangential discontinuities giving allowance to violate frozen flux physics (representing the decoupling between plasma and field lines). Besides, current sheets favor fast current dissipation, thereby facilitate the fast topological rearrangement of magnetic field lines \citep{Dungey01071953}. This process, known as magnetic reconnection, was profoundly advanced and established by the seminal work of \citet{1957JGR....62..509P}. During magnetic reconnection, the dissipation of current sheet converts magnetic free energy into plasma in forms of thermal energy, kinetic energy, and particle acceleration. Current sheets exist across an extremely wide range of magnetized plasma systems, which span vastly different physical parameters including spatial scales, ionization degrees, temperatures, densities, magnetic field strengths, and collisionality \citep{2010RvMP...82..603Y,2020RSPSA.47690867N,1989ApJ...340..550Z}, ranging from small-scale laboratory plasmas to planet magnetotails, collisionless interplanetary space, fully-ionized high-temperature solar corona, and even weakly-ionized low-temperature solar chromopshere. Therefore, physics of current sheets is crucial for understanding various magnetically explosive activities or nonideal magnetohydrodynamics processes in magnetized plasma systems.

Solar atmosphere provides a natural magnetized plasma laboratory for studying magnetic reconnection processes. It hosts a variety of such events, which span a wide range of scales: from large-scale solar flares, coronal jets, and large-scale magnetic flux rope (MFR) eruptions \citep{2011LRSP....8....6S,2016NatCo...711522J}; to mid-scale microflares, transition-region jets, and micro-MFR eruptions \citep{2015NatCo...6.7008W,2023NatCo..14.2107C}; and down to small-scale events such as ultraviolet bursts (UVBs), Ellerman bombs, chromospheric jets, and chromospheric reconnection \citep{2013ApJ...769L..33Z,2016ApJ...824...96T, 2016ApJ...823..110R,2014Sci...346C.315P, 2014Sci...346A.315T,2012ApJ...759...33S, 2021A&A...646A..88N, 2023A&A...673A..11R}. Durations of these reconnection events are typically last less than a few hours, suggesting rapid dissipation of currents in the reconnection current sheets within the solar atmosphere. Lundquist number, a dimensionless number defined as the ratio of Alfvén wave timescale to resistive diffusion timescale, is a indicator describing dissipated rate of current sheets. Once reconnection enters the tearing instability stage, Lundquist number scales weakly on dissipated rate of current sheets. The reconnection rate, also named Alfvén Mach number ($M_A$)—defined as the rate of magnetic flux conversion and applicable to both linear and nonlinear reconnection stages—provides a more fundamental description of the dissipation rate of current sheets \citep{2010PhRvL.105w5002U}. Recent observational studies show reconnection rates $M_A$ of order of $10^{-2}$ to $10^{-1}$ for solar atmospheric reconnection events \citep{2001ApJ...546L..69Y,2013NatPh...9..489S,2012ApJ...745L...6T,2022NatCo..13..640Y}. This suggests that solar atmospheric reconnection events are in form of fast reconnection, proceeding at rates much higher than those predicted by the Sweet-Parker model dominated by Ohmic dissipation ($M_A$$\sim$$10^{-7}$). Despite the landmark contribution of the Sweet-Parker model to our understanding of magnetic reconnection, its predicted rate is far too slow to explain observed solar reconnection events. The search for dissipation mechanisms that can effectively speed up Sweet-Parker reconnection is an important part of the history of reconnection theory development. Historically, slow-mode shocks that rely on anomalous resistivity were introduced to speed up reconnection process, namely Petschek reconnection \citep{1964NASSP..50..425P}. A widely accepted view is that magnetic reconnection operating in different layers of the solar atmosphere requires different dissipation mechanisms. In the solar corona, Hall effect is one of fast dissipation mechanisms. In the solar chromosphere, however, neutral collisions are so negligible that partial ionization effect (i.e., ambipolar diffusion) becomes a necessary consideration \citep{1994ApJ...427L..91B,2022NatRP...4..263J}. Despite different physical environments in different layers, the tearing instability of reconnecting current sheets is a common dissipation mechanism bridging reconnection processes in vast range of physical environments. Tearing instability is triggered when ratio of the current sheet length to width is large enough, with critical Lundquist number of $10^4$. Consequently, resulting plasmoids by tearing instability can speed up reconnection process greatly \citep{2009PhPl...16k2102B,2001EP&S...53..473S,2022NatCo..13..640Y}. Recent studies evidence that tearing instability operates in multi-scale reconnection events of different layers in solar atmosphere.

The formation of a current sheet is often a key outcome of the driven process that leads to magnetically eruptive activities, and thereby studying of current sheet forming holds a key to understand causalities of these erupting activities. As suggested by theoretic studies, the continuous increase of magnetic stress at tangential discontinuities leads to the formation of current sheets \citep{1971JETP...33..933S,1996PhPl....3.2885L}, and magnetic reconnection is a process that tries to decrease increasing magnetic stress. Solar atmosphere hosts a variety of activities that enhance magnetic stress, suggesting a diversity of driven processes for current sheet formation. A common driven process of current sheet formation is the eruption of MFRs. As an MFR moves outward, it stretches the surrounding field lines, thereby increasing the magnetic stress behind it which facilitates the formation of a current sheet \citep{2005ApJ...622.1251L,2022ApJ...933..148C,2000JGR...105.2375L,2016NatCo...711522J,2001EP&S...53..473S,2018ApJ...853L..18Y,1995ApJ...451L..83S,1992PASJ...44L..63T,1998ApJ...499..934O,1994Natur.371..495M}. Another common driver for increasing magnetic stress are photospheric flows, such as shearing and converging motions. These motions bring oppositely directed field lines into close proximity, which amplifies the magnetic stress at the tangential discontinuities and ultimately leads to current sheet formation \citep{2018ApJ...862L..24P,2020ApJ...891...52S,2019ApJ...872...32S}. Such photospheric motions have been evidenced as drivers of many kinds of reconnection activities, including tether cutting reconnection \citep{2001ApJ...552..833M,2021NatAs...5.1126J,2015NatCo...6.7008W,2002JGRA..107.1164T}, cancellation nanoflares \citep{2018ApJ...862L..24P,2020ApJ...891...52S,2019ApJ...872...32S}, UVBs \citep{2014Sci...346C.315P}, chromospheric reconnection \citep{2024ApJ...966L..29C,2017ApJ...839...22H}, braiding reconnection and nanoflares \citep{1988ApJ...330..474P,1983ApJ...264..642P}.

Beyond the aforementioned processes, a third primary driver for current sheet formation is the emergence of magnetic flux through the photosphere. This process is essential for transporting magnetic flux from the solar convection zone and into the solar atmosphere. As emerging magnetic flux enters the solar atmosphere, it inevitably compresses the overlying magnetic field lines in the atmosphere. This process increases magnetic stress and can lead to the formation of reconnecting current sheets at magnetic tangential discontinuities, often referred to as the emerging-reconnection process. \citet{1976SoPh...48...89T} first established the concept of the emerging flux reconnection process with an analytical model. \citet{1977ApJ...216..123H} further linked this process to various solar magnetic activities. Since then, the model has been significantly advanced and is now widely used to explain the formation of solar jets, including bidirectional jets, as well as coronal and chromospheric anemone jets \citep{1994ApJ...431L..51S,1995Natur.375...42Y,2008ApJ...683L..83N,2025ApJ...987..148W}. Recent observational and simulation studies suggest that emerging-reconnection processes also contribute significantly to solar atmosphere heating \citep{2024NatAs...8..706L,2025ApJ...985..152Y}. Moreover, emerging-reconnection processes have deep influence on erupting processes of MFRs. \citet{2000ApJ...545..524C} showed that the emerging-reconnection process plays a dual role in MFR eruptions, acting as both a facilitator and an inhibitor. 


Earlier simulation studies demonstrate a physical scenario that emerging fluxes can successively create and elongate reconnecting current sheets, as well as trigger their tearing (see \citet{1995Natur.375...42Y,2024NatAs...8..706L,2017ApJ...841...27N,2024A&A...685A...2C}). These findings indicate that emerging fluxes play a crucial role not only in the formation of reconnecting current sheets but also in modulating their dissipation. We present an observational study on the role of photospheric emerging fluxes in the formation and subsequent development of a reconnecting current sheet in the corona. Motivated by these findings, this paper presents an observational study addressing how photospheric emerging fluxes drive the formation and subsequent evolution of a coronal current sheet. The observed evolution of the current sheet closely matches the findings of previous numerical studies, validating the theoretical models. In addition, we not only identify two key heating processes of the current sheet, but also demonstrate how plasmoids cause fast magnetic reconnection. Section 2 describes the observational data and methodology, Section 3 and 4 present the detailed observational results, and Section 5 provides our conclusions and discussions.

\section{Instruments and Data}

This event was observed by the space-based Solar Dynamics Observatory (SDO). Extreme ultraviolet (EUV) observations from the Atmospheric Imaging Assembly \citep[AIA;][]{2012SoPh..275...17L} and photospheric line-of-sight (LOS) magnetograms from the Helioseismic and Magnetic Imager \citep[HMI;][]{2012SoPh..275..229S} are utilized. The SDO/AIA images have a cadence and pixel size of 12 s and 0\arcsec.6, respectively, while the SDO/HMI magnetograms have a cadence of 45 s and a pixel size of 0\arcsec.5. The data were processed using the standard calibration procedures via the Solar Soft Ware (SSW). The reconnection event, which lasted for approximately two hours, occurred near the solar equator on the eastern limb on 26 October 2022, with approximate heliocentric coordinates of [-920\arcsec, -220\arcsec]. For consistency with the conventional observational perspective that solar jets are directed upwards, all images presented in this paper have been rotated clockwise by 90°. Furthermore, to facilitate temperature analysis, we derived differential emission measure (DEM) maps for the region of interest using six of SDO/AIA's extreme ultraviolet (EUV) filters (94, 131, 171, 193, 211, and 335 Å) and computed the resulting average temperature maps {using the ``xrt\_dem\_iterative2.pro" routine in SSW package. The DEM is determined by the following equation:}

\begin{equation}
	{I_i}=\int{R_i}(T){\times}\rm{DEM}(\emph{T})d\emph{T},
\end{equation}
{where ${I_i}$ represents the observed emission intensity of an AIA wavelength channel $i$; ${R_i}(T)$ stands for the temperature response function of the channel $i$; DEM($T$) is the DEM of coronal plasma. After solving the DEM, one can obtain the Emission Measure (EM) within a temperature range ($T_{min}$, $T_{max}$) through the following equation:}

\begin{equation}
	\rm{EM}=\mathit{\int_{T_{min}}^{T_{max}}}DEM(\emph{T})d\emph{T}.
\end{equation}	
{Finally, one can estimate the average temperature through EM and DEM:}
\begin{equation}
	\bar{T}=\frac{\int_{T_{min}}^{T_{max}}\mathrm{DEM}(T){\times}T\mathrm{d}T}{\rm{EM}},
\end{equation}
{where $\bar{T}$ is the weighted average temperature \citep{2012ApJ...761...62C}. It should be noted that the derived $\bar{T}$ should be regarded as a diagnostic proxy rather than a measure of the true plasma temperature.}

{The average temperature maps derived from Equation (3) do not include temperature errors. For cases requiring higher accuracy, DEM curves are supplemented. These DEM curves are obtained via Monte Carlo (MC) simulations, incorporating observational errors introduced by ``aia\_bp\_setimate\_error.pro" routine in SSW package (see \citet{2012ApJ...761...62C} for details). The resulting distribution of DEM solutions defines the uncertainty, represented by its scattering area. We calculate the average temperature, $\bar{T}$, and its associated uncertainty directly from the best-fit DEM solution and the scattering area of MC simulations, respectively. }

\section{Results} 
\subsection{Responses in the layered atmosphere}

The region of interest (ROI) is an ephemeral region. Figure 1 shows the temporal evolution of the (ROI) in the layered solar atmosphere, with its field of view (FOV) of $68\arcsec \times 115\arcsec$. The panels from top to bottom display observations from the SDO/HMI, SDO/AIA 171~\AA and SDO/AIA 131~\AA. SDO/HMI observations map LOS magnetic field at source region. The 171~\AA\ channel is sensitive to plasma at the temperature $10^{5.8}$~K, while the SDO/AIA 94~\AA channel sensitive to plasma at temperatures $10^{5.6}$ and $10^{7}$~K. This combination of multi-wavelength observations comprehensively diagnoses plasma dynamics of the ROI at various temperatures and the evolution of magnetic fields. The temporal evolution of this reconnection event proceeds from left to right. We divide the event into two stages-- lengthening (19:00 UT-20:11 UT, characterized by the current sheet forming, lengthening, and tearing with low frequency) and shortening (20:11 UT-22:00 UT, characterized by the current sheet shortening and tearing with high frequency) phases--based on the change in the current sheet length. {The term ``current sheet'' typically describes dissipation regions of reconnection events, but given the fact that magnetic reconnection dissipation occurs at microscopic plasma scales \citep{1997aspp.book.....T}, we will henceforth refer to the macroscopic observational manifestation of the observed reconnection process as the ``plasma sheet" rather than the ``current sheet."}

In SDO/HMI images (Figure 1a-e), black and white areas respectively represent negative and positive magnetic polarities, while grey areas correspond to photospheric surroundings. As revealed by Figure 1a, the source region was located in a mixed-polarity region. Three key magnetic polarities that are related to this reconnection event are labeled P1, P2, and N1 (see Figure~1a), with N1 sandwiched between P1 and P2. It can be seen that areas of N1 and P2 increase throughout the event, indicating the continuous emergence of their magnetic fluxes. For quantitative analysis, we measure magnetic fluxes of P1 and N2 in each panel of Figure 1a-e, and label the values at the top left. The measurements show that both negative and positive magnetic fluxes increase primarily in the lengthening stage, while they increase only slightly in the shortening stage. Specially, the positive flux (P1) and the negative flux (N2) increased by just 0.7$\times$10$^{19}$ and 0.2$\times$10$^{19}$ Mx, respectively, in the latter stage.

The SDO/AIA 171 \AA\ channel reveals no significant magnetic structures or magnetic activities initially in the source region at 19:00 UT (see Figure 1f). At 19:53 UT, however, a reconnecting {plasma sheet} appeared above the mixed-polarity region as the negative flux of N1 increased significantly ($\vert\delta\phi_{N1}\vert$=2$\times$10$^{19}$ Mx). The reconnection region consists of two cusps with a curved {plasma sheet} between them, with the lower cusp and one bifurcation of the upper cusp rooted back in the solar surface and the other bifurcation of the upper cusp forming a solar jet that extends into the higher corona (see Figure 1g). To directly compare the magnetic connectivity of the reconnection region in the atmosphere with the unverlying mixed polarities in the photosphere, we created a composite image (Figure 1b) by superimposing the SDO/AIA 171 \AA\ map at 19:53 UT onto the corresponding SDO/HMI map. Figure 1b shows that the lower cusp was rooted in N1 and P1, representing the post-reconnection loop, while the leg of the upper cusp was rooted in P2. This magnetic topology is consistent with the anemone jet model \citep{1994ApJ...431L..51S,1995Natur.375...42Y,2008ApJ...683L..83N,2025ApJ...987..148W,2007Sci...318.1591S}, and indicates the occurrence of reconnection between close loop N1P2 with ambient field lines rooted in P1. Given significant emerging of polarity N1 at this time, we infer that close loop N1P2 originated from magnetic flux emerging in the photosphere. Consequently, the observed reconnecting {plasma sheet} formed as a result of reconnection between this emerging loop (N1P2) with the ambient field lines rooted in P1. Note that a slight decline instead of emergence in the positive magnetic flux of P2 was detected during this interval. We infer that this transient change may stem from spatial averaging with the co-located negative flux of N1, since the two polarities were still in close proximity and not highly separated (Figure 1b).

The lower and upper cusps further separated after the formation of the reconnecting {plasma sheet} (see Figure 1h), indicating its further lengthening. As the {plasma sheet} lengthened, three well spatially resolved plasmoids appeared one-by-one within it, indicating the onset of tearing instability. Note that the left brightening point shown in the subgraph of Figure 1h is the apex of post-reconnected loops rather than a plasmoid. Observational facts suggest that the onset of tearing instability is closely related to the lengthening of the {plasma sheet}, consistent with the theoretical model \citep{1963PhFl....6..459F}. The {plasma sheet} reached its maximum length at 20:11 UT, after which it entered the shortening phase (see Figure 1i and j).

Observational features revealed by the SDO/AIA 131 \AA\ are highly similar to that in the SDO/AIA 171 \AA\, including two cusps, the lengthening and shortening of the {plasma sheet}, as well as plasmoids (see Figure 1l--o). This similarity is due to the proximity between one of the sensitive temperatures of the SDO/AIA 131 \AA\ ($10^{5.6}$ K) and the primary response temperature of SDO/AIA 171 \AA\ ($10^{5.8}$ K). But fine differences emerge from a close comparison of the SDO/AIA 131 Å and 171 Å channels: prominent loops underlying the lower cusp (see Figure 1o). These loops represent the post-reconnection loops created by the reconnection process. Their discordant visibilities in the SDO/AIA 131 \AA\ and 171 \AA\ channels indicate their high temperature, possibly corresponding to the high temperature component of SDO/AIA 131 \AA\ response temperatures ($10^{7}$ K).

\subsection{Development of the reconnecting {plasma sheet}}

To supplement the qualitative findings from the previous section, we utilize integrated curves and space-time slices to quantitatively analyze the relationships among the evolution of photospheric magnetic fluxes, photospheric footpoint motions, as well as the physical parameters (length, height, temperature) of the {plasma sheet}, as shown in Figure 2. Figure 2a shows the integrated magnetic flux profiles for P1 and N2, and a space-time slice of their motion (S1), with the integration area and slice path indicated in Figure 2f. The negative flux (N2) exhibited an impulsive rise starting at 19:30 UT, followed by a gradual increase until the end, while the positive flux (P1) increased gradually from the beginning, suggesting emerging of magnetic fluxes (see Figure 2a). Concurrently, N1 and P2 showed divergent motion (see Figure 2a), further strengthening the flux emerging scenario \citep{1999ApJ...527..435S,2010ApJ...720..233C,2007A&A...467..703C,2008ApJ...687.1373C,2020ApJ...900...84W}. These results provide support for the emerging reconnection process discussed in the previous section.

During the reconnection process, the {plasma sheet} height continually rose instead of remaining constant, as shown by the EUV observations in Figure 1. To quantitatively analyze the height variation of the {plasma sheet}, we constructed a space-time slice (S2) with the slice path crossing it, as shown in Figure 2b. The rise of the {plasma sheet} displayed two distinct phases, separated by a transition at 20:11 UT, with velocities of 2.56 km/s and 0.63 km/s before and after the transition, respectively (see Figure 2b). 

Besides the height of the {plasma sheet}, its length also underwent significant changes. To display these changes in detail, we created a space-time slice (S3) with the slice path along the {plasma sheet}, as shown in Figure 2c. S3 reveals the existence of two brightening boundaries, which essentially correspond to the lower and upper cusps of the reconnection region, and the region between the two cusps thus represents the reconnection {plasma sheet}. The two cusps first diverged and then began to converge, suggesting an initial lengthening followed by a shortening of the {plasma sheet}, as also seen in Figure 1. During the lengthening phase (19:40 UT--20:11 UT), the upward and downward elongation speeds of the {plasma sheet} were 6.2 km/s and 0.76 km/s, respectively. The downward elongation speed was much smaller than the upward speed, and this asymmetry can be attributed to the downward propagation occurring in a lower layer where the gas and magnetic pressure were significantly higher. When entering the shortening phase (after 20:11 UT), the lower and upper cusps approached at speeds of 0.8 and 0.9 km/s, respectively. It is particularly noteworthy that a transition separating the lengthening and shortening phases also occurred at 20:11 UT, coincide with the transition in Figure 2b.

To analyze the temperature evolution of the {plasma sheet}, we created a space-time slice (S4) of the averaged temperature maps along the same path as S3 (Figure 2d). Two significant features emerge from S4. One is that the lower cusp maintains a high temperature (around $10^{6.8}$ K) throughout the entire reconnection event, noticeably higher than the upper cusp ($10^{6.2}$--$10^{6.6}$ K). The other is the temporal distribution of the {plasma sheet} temperature: a low-temperature phase (19:40 UT--20:11 UT) and a subsequent high-temperature phase (after 20:11 UT). Two temperature phases are also separated by a transition at 20:11 UT.

Close inspection of S3 reveals the ubiquity of brightening stripes in the {plasma sheet, and they appear as locally brightening areas propagating along the plasma sheet in animated Figure 1. The measured peak propagation speed is 580 km/s. Given the relatively low 12s cadence of AIA observations and projection effects in the plane of sky, this value likely underestimates their true maximum speed. An alternative interpretation could relate these stripes to propagating bright dots observed in coronal plumes \citep{2025ApJ...988..133W}. However, such dots typically propagate at tens of km/s \citep[comparable to the coronal sound speed;][]{2025ApJ...988..133W}, markedly slower than observed stripes that propagate at hundreds of km/s (comparable to the Alfvén speed in the corona). Besides significant difference in speeds, another significant distinction stems from their underlying magnetic polarities: the propagating dots originate from a predominantly unipolar region \citep{2025ApJ...988..133W}, whereas the observed stripes emerge from a mixed-polarity region (see Figure 1). These two differences suggest distinct origins. Here we infer that the observed stripes are more likely magnetically driven plasmoids according to their propagating speeds comparable to a coronal Alfvén speed and a mixed-polarity origin.} The presence, direction and slope of each brightening stripe thus represent the formation, propagation direction and propagation speed of a plasmoid. We applied a low-pass filter to S3 to enhance brightening stripes, with the result shown in Figure 2e. The filtered result reveals that the brightening stripes appear in two distinct phases: a low-frequency phase ($\sim$ {8 mHz}; 19:40–20:11 UT) and a subsequent high-frequency phase ($\sim$ {15 mHz}; after 20:11 UT). Two frequency phases are also separated by a transition at 20:11 UT.

The transitions of the rising speed, length and temperature of the {plasma sheet} as well as the plasmoid frequency exhibit coincidence, as shown by the comparison of Figure 2b--e. This coincidence points to a common underlying driver, which we will explore in a later section.

 Another noteworthy point is plasmoid heating effects, an example of which is demonstrated by Figure 2g and h (zoom-in versions of Figure 2c and d). The collision between upward-propagating plasmoids and the upper cusp is clearly seen in Figure 2g. {Collisions of plasmoids cause} local heating (see Figure 2h), with a temperature {in a range of $10^{6.4}$--$10^{6.6}$ K}. This clearly illustrates the heating process of plasmoids in the {plasma sheet}. A detailed analysis of heating process will be presented in the next section.

\subsection{Case-by-case analysis of plasmoid heating effects}

The comparison of Figures 2d and 2e suggests that the heating of the {plasma sheet} is closely related to plasmoids, with one case exemplified in Figure 2g and 2h. Close inspection of the present event throughout reveals four distinct heating processes in the {plasma sheet}, including the plasmoid-plasmoid coalescence (Figure 3a), plasmoid-lower-cusp collision (Figure 3b), plasmoid-upper-cusp collision (Figure 3c) and tearing of the {plasma sheet} (Figure 3d). In this section, we provide a case-by-case analysis of the four distinct heating processes to elucidate their underlying physical mechanisms.

\textbf{Plasmoid-plasmoid coalescence}. At 19:38:18 UT, Figure 3a1 reveals the existence of two plasmoids in close proximity within the {plasma sheet}, which were significantly brighter than their surroundings. Their morphology and spatial separation are further clarified in the corresponding temperature map, which displays the structures with sharper edges (see Figure 3a2). Their peak temperatures reach up to $10^{6.65}$ K (see Figure 3a2). A half minute later, a larger plasmoid was formed as a result of the convergence and coalescence of two spatially-separated plasmoids (Figure 3a3). Besides of a larger spatial scale, the peak temperature of the newly-forming plasmoid increased to $10^{6.95}$ K, indicating the heating effect of the plasmoid-plasmoid coalescence (Figure 3a4). This heating effect is likely attributed to the secondary {plasma sheet} formed between converging plasmoids as predicted by theoretical studies \citep{1977PhFl...20...72F,1979PhFl...22.2140P,2011ApJ...733..107K,2012A&A...541A..86K}. Secondary {plasma sheets} play a crucial role not only in the local heating, but also in the merging process of plasmoids.

\textbf{Plasmoid escaping downward}.
Once formed, plasmoids rapidly escape from the {plasma sheet}, as evidenced by the stripes in Figure 2e. Some of these plasmoids propagate towards the lower cusp; we use Figure 3b to illustrate this phenomenon in detail. Before escaping at 20:00:57 UT, the plasmoids manifested as faint, brightening blobs (Figure 3b1). They correspond to two high-temperature regions (peak temperature $\sim$ $10^{6.61}$ K) with sharper edges, as shown in Figure 3b2. About one minute later, the plasmoids escaped downward and collided with the lower cusp. The heating effect is manifested as a prominent enhancement in both the lower cusp's intensity at the SDO/AIA 171 Å channel (Figure 3b3) and its temperature (peak temperature $\sim$ $10^{6.84}$ K; Figure 3b4) following the collision, similar to the heating of post-flare loops. The heating effect revealed in Figure 3b can be attributed to diverse mechanisms, including the formation of secondary {plasma sheet} formed between the lower cusp and propagating-downward plasmoids \citep{2010ApJ...713.1292M,2008A&A...477..649B}, as well as fast-mode shocks \citep{2013ApJ...767..168L,2024ApJ...971...85C,1994Natur.371..495M,1983SoPh...84..169F,2012ApJ...753...28G,2025ApJ...982..142R}.

\textbf{Plasmoid escaping upward}.
In this event, most of plasmoids escaped towards the upper cusp (see Figure 2e); we use Figure 3c to illustrate this phenomenon in detail.At 20:27:45 UT, an ellipse-shaped plasmoid appeared in the middle of the {plasma sheet} (see Figure 3c1 and 3c2). Twelve seconds later, a smaller plasmoid formed from the former plasmoid due to tearing, manifesting as a brightening blob in the SDO/AIA 171 \AA\ channel (Figure 3c3) and a high-temperature blob in the temperature map (Figure 3c4). In the following half minute, the newly formed plasmoid escaped upward and collided with the upper cusp (Figure 3c5, c6). The collision led to its coalescence with the cusp and triggered emission enhancement (Figure 3c5) as well as significant local heating ($10^{6.93}$ K; Figure 3c6). This case demonstrates a heating effect, responsible for the intermittent heating in Figure 2g and h, caused by a plasmoid-cusp collision, analogous to the case presented in Figure 3b. Another noteworthy point is that shortly after the upper cusp heated, a plasma blob detached from it and propagated along the collimated jet. The blob appeared as a region brighter than the surrounding jet background (Figure 3c7), with a temperature significantly higher than its environment ($10^{6.82}$ K; Figure 3c8). More importantly, the temperature map revealed that the blob was detaching from the upper cusp, indicating its origin there. The secondary {plasma sheet} formed between the propagating-upward plasmoid and the upper cusp is a potential mechanism not only for their coalescence but also for the subsequent heating effect and the formation of the hot blob in the jet \citep{2017ApJ...841...27N}. Plasmoid-driven shocks are an alternative heating mechanism \citep{2017ApJ...841...27N}.

\textbf{The tearing of the {plasma sheet}}.
Besides the three heating processes mentioned above, we now introduce the fourth: {plasma sheet} tearing, as shown in Figure 3d. At the beginning of this case (20:06:21 UT; Figure 3d1), the {plasma sheet} appeared as a faint elongated structure at a temperature of about $10^{6.59}$ K (Figure 3d2). About three minutes later, this {plasma sheet} entered the tearing phase, as indicated by the appearance of well-resolved plasmoids within it (see Figure 3d3). These plasmoids manifested as a series of elliptical bright blobs in the SDO/AIA 171 Å channel (see Figure 3d3). They appeared as high-temperature blobs with a more distinct profile ($10^{6.88}$ K) in the corresponding temperature map and were separated by secondary X points (see Figure 3d4). About one minute later, the plasmoids increased significantly in size (see Figure 3d5) and temperature ($10^{6.93}$ K; Figure 3d6). At the same time, a small plasmoid escaped upward, then collided with the upper cusp {and resulted in heating at the cusp} (see Figure 3d7 and d8), closely similar to the case in Figure 3c. Such similarity indicates that they share the same heating mechanism.

The case in Figure 3d clearly demonstrates the sustained heating process of the {plasma sheet} represented by appearances of high-temperature plasmoids originating from the tearing process. This heating process indeed is related to the internal heating of the {plasma sheet} in Figure 2d. Two underlying findings worth paying particular attention to are: (1) plasmoids occurred and the {plasma sheet} was heated within a short time interval of a few minutes, indicating the high efficiency of both the tearing and resulting heating processes; (2) the {plasma sheet} temperature increased with plasmoid size, indicating a direct correlation between the two. These two underlying findings have significant implications for the heating mechanism of the {plasma sheet} via its tearing, and are crucial for a deeper understanding of the process. The details will be presented in the next section.

\subsection{Analytical Model of {Plasma Sheet} Tearing}

The {plasma sheet} did not maintain a consistently high temperature; rather, it was heated intermittently by its tearing processes. The tearing process proceeded rapidly and heated the sheet efficiently, as indicated by its short time scale (see Figure 3d). In this section, we will elucidate their underlying physical significance and their determinant through an integrated approach of observational data and analytical methods. The analytical analysis defaults to focusing on the two-dimensional magnetic reconnection process.

\subsubsection{Reconnection rate induced by {plasma sheet} tearing}

The tearing process of a {plasma sheet}, namely the tearing instability, is triggered as a result of the continual increase  of the tearing wavenumber, which linearly grows with $S^{3/8}$ \citep{2009PhRvL.103j5004S,2017ApJ...849...75H}, where $S$ is the Lundquist number. Once the tearing instability is triggered, secondary {plasma sheets} form and tear the main {plasma sheet}, and as a result, plasmoids are formed by secondary reconnection. We will next analyze the magnetic flux conversion during the tearing process and derive its analytical description.

During the tearing process of the main {plasma sheet}, the injection rate of magnetic fluxes in the reconnection inflow region is presented by ${\partial \phi_{inflow}}/{\partial t}$ {(see Appendix for details)}, in which

	\begin{equation}
	\frac{\partial \phi_{inflow}}{\partial t} = {v_{in}}B_0  +  \eta_m \nabla^2\phi,
\end{equation}
where $v_{in}$, $B_0$ and $\phi_{inflow}$ are the inflow speed, magnetic field strength and magnetic fluxes in the reconnection inflow region. 


{Equation (4) indicates two dependent dissipation processes of ${\partial \phi_{inflow}}/{\partial t}$}. One is $\eta_m \nabla^2\phi$, representing the conversion rate of magnetic fluxes by the Ohm dissipation that relies on the magnetic diffusivity $\eta_m$, equal to $\phi_{\eta_m}/\delta t$ \citep{1957JGR....62..509P}. The other is ${v_{in}}B_0$, {attributed to} plasmoid widening process, during which plasmoids grow wider by consuming magnetic fluxes converted from the inflow region \citep{2010PhRvL.105w5002U}. Assumed that $N$ plasmoids of comparable width are generated by the tearing of the main {plasma sheet}. The time from the plasmoid's birth until it reaches its maximum width scale is denoted as $\tau_{grow}$, with its width growing from 0 to $w_p$. The magnetic fluxes consumed by the widening of a plasmoid are $\phi_p$=$w_pB_0$. Therefore, the magnetic flux growth rate caused by the widening of a plasmoid is given by:
	\begin{equation}
	\frac{\phi_{p}}{\tau_{grow}}=\frac{w_pB_0}{\tau_{grow}},
\end{equation}
where $\frac{w_p}{\tau_{grow}}$ represents the widening speed of plasmoids, namely the component of the plasmoid enlargement speed perpendicular to the main {plasma sheet} (hereafter denoted as $v_{p\perp}$). Therefore, Equation (5) is simplified to 
	\begin{equation}
	\frac{\phi_{p}}{\tau_{grow}}={v_{p\perp} B_0}.
\end{equation}

During the tearing of the {plasma sheet}, magnetic flux must conserve. Therefore, the magnetic flux injection rate in the inflow region (${\partial \phi_{inflow}}/{\partial t}$) equals the sum of the magnetic flux growth rate consumed by $N$ magnetic plasmoids ($N\phi_p/\tau_{grow}$) and the magnetic flux conversion rate due to Ohmic dissipation ($\phi_{\eta_m}/\delta t$). However, given the extremely low efficiency of magnetic flux conversion via Ohmic dissipation (corresponding to a reconnection rate of only $10^{-7}$), it can be neglected. Thus, the approximate expression ${\partial \phi_{inflow}}/{\partial t}$$\sim$$N\phi_{p}/\tau_{grow}$ holds. By combining Equations (4) and (6) {and neglecting the Ohmic dissipation term ($\eta_m \nabla^2\phi$)}, we obtain
	\begin{equation}
	v_{in}\sim Nv_{p\perp}.
\end{equation}
From Equation (7), the expression for the Alfvén Mach number (reconnection rate) during {plasma sheet} tearing can be derived:
	\begin{equation}
	M_A=	\frac{v_{in}}{v_A}\sim \frac{Nv_{p\perp}}{v_A},
\end{equation}
where $v_A$ represents the Alfvén speed in the inflow region. This expression suggests $v_{p\perp}$, $N$, and $v_A$ as observational indicators for the reconnection rate of a torn {plasma sheet}, which can also be derived from the analytical work of \citet{2010PhRvL.105w5002U}. Since the maximum velocity of magnetic plasmoids represents the local Alfvén speed \citep{2008A&A...477..649B}, we take $v_A$ to be the maximum observed propagation speed of the magnetic plasmoid in Figure 2d, namely $v_A$ = 580 km/s. Apart from $v_A$, both $N$ and $v_{p\perp}$ also are measurable quantities. Next we will combine Equation (8) with observational data to calculate the reconnection rate during the tearing of the {plasma sheet}.

Throughout this reconnection event, the {plasma sheet} underwent multiple instances of tearing, each of which generated new plasmoids. However, only one plasmoid showed a clearly measurable widening process due to the limitation of the spatial resolution of the SDO/AIA. The plasmoid widening was measured from the slice in Figure 4b, whose path was oriented perpendicular to the {plasma sheet}, as shown in Figure 4a. The plasmoid widening bi-directionally from {20:09:21 UT to 20:09:57 UT} (see Figure 4b). {The projected speeds in each direction, denoted $v1$ and $v2$, sum to a total perpendicular speed of $v_{p\perp}$ = 69 $\pm$ 13 km/s. The quoted uncertainty is the standard deviation from 10 independent measurements.} The number of distinguishable plasmoids $N$ equals {1} (see Figure 4a and Figure 4c). Substituting the measured values of $v_{p\perp}$, $N$, and $v_A$ into Equation (8) yields an Alfvén Mach number of $M_A \approx 0.12\pm0.02$. This $M_A$ represents the rate of secondary reconnection introduced by the tearing process. This calculated order of $M_A$ indicates that secondary reconnection lies in the fast reconnectio regime, which enables the rapid dissipation of magnetic field. Accordingly, plasmoid formation and plasma heating—direct products of the secondary reconnection—occur rapidly, accounting for their short time scale witnessed in Figure 3d. However, it should be noted that the observations represent the projection of three-dimensional structures onto a two-dimensional plane, while the calculated values are approximations derived under a two-dimensional framework and with Ohmic dissipation neglected. {Moreover, $N$ corresponds only to the number of well-resolved plasmoids and is therefore likely an underestimate, as indicated by comparing with plasmoids in simulations \citep[i.e.,][]{2025ApJ...993...31D}. Furthermore, the plasmoid widening occurs on a timescale too short to be well resolved by the 12-second cadence of AIA EUV observations}. Given these limitations as well as the approximations in the model, the calculated reconnection rate $M_A$ can not fully represent the actual physical conditions observed. The purpose of estimating the reconnection rate here is to provide an order-of-magnitude reference, aiding in understanding the enhancement of reconnection rate by the the {plasma sheet} tearing.

\subsubsection{Energy flux induced by {plasma sheet} tearing }

Figure 3d reveals a positive correlation between the main {plasma sheet} temperature and plasmoid width. To understand this correlation, we now employ an analytical investigation to elucidate the underlying physics. During the tearing of the main {plasma sheet}, the energy release rate of secondary {plasma sheets} is proportional to the Poynting flux $S$ in the reconnection inflow region, which is given by:
	\begin{equation}
	S \propto \frac{v_{in}B^2_0}{4\pi},
\end{equation}
where $B_0$ and $v_{in}$ are the magnetic field strength and speed in the reconnection inflow region. Substituting Equation (7) into (9) and multiplying the result by $\tau_{grow}$ yields the released energy $E$ during the generation of $N$ plasmoids within time $\tau_{grow}$:
	\begin{equation}
	E  = S\tau_{grow} \propto \frac{Nv_{p\perp} B^2_0 \tau_{grow}}{4\pi} = \frac{Nw_{p} B^2_0 }{4\pi}.
\end{equation}
This equation implies that the energy released by the secondary reconnection, which is caused by the tearing of the main {plasma sheet}, can be quantified by observational measurements of the plasmoid number $N$ and width $w_p$. This equation comes to a conclusion that for a fixed number of plasmoids $N$, the released energy monotonically scales with the plasmoid width, which is testable. To test this conclusion, we compare the temperature evolution of plasmoids in different widths, as shown in Figure 4c and 4d. The comparison clearly shows that the peak temperature of plasmoids increased as they widened, providing support for Equation (10). Additional support can also be seen from the case in Figure 3d, consistent with the prediction of Equation (10).

\subsection{Analytical Model of Plasmoid Ejection}

\subsubsection{Reconnection rate induced by plasmoid ejection}

Ejections of plasmoids also have deep influence on the heating process in a torn {plasma sheet}. Figure 4e and 4f show the evolution of the heated area of the main {plasma sheet} after the plasmoids shown in Figure 4c and 4d were ejected. It is observed that the heated area of the main {plasma sheet} increased substantially concurrent with these ejections (Figure 4e, 4f), within a time interval on the order of only minutes. Such a short timescale indicates that plasmoid ejections play a key role in accelerating magnetic reconnection, which in turn is responsible for the rapid heating of the {plasma sheet} \citep{2022NatCo..13..640Y,2009PhPl...16k2102B,2010PhRvL.105w5002U,2017ApJ...849...75H}. In this section, we will derive the expression for the reconnection rate induced by plasmoid ejections, and elucidate its underlying physical significance through an integrated approach of observational data and analytical methods. Our analytical treatment defaults to focusing on the two-dimensional magnetic reconnection process.

Considering the conservation of magnetic flux between the injection rate in the inflow region and the ejection rate in the outflow region, we have ${\partial \phi_{inflow}}/{\partial t}$=$\phi_{eject}/\delta t$, where $\phi_{eject}$ and $\delta t$ represent the newly converted magnetic fluxes in the outflow region and their conversion time, respectively. We assume that the flux ejection rate is the sum of two components: 
\begin{equation}
	\frac{\phi_{eject}}{\delta t}=\frac{\phi_{\eta_m}}{\delta t} + \frac{\phi_p}{\tau_{eject}},
\end{equation}
where $\phi_{\eta_m}/\delta t$ is the rate due to Ohmic dissipation, and $\phi_p/\tau_{eject}$ is the flux transport rate caused by plasmoid ejections; ${\phi_p}$ and $\tau_{eject}$ are the magnetic flux of a plasmoid and its ejected timescale, respectively. We have $\phi_p\sim w_p B_0$, where $w_p$ and $B_0$ are its width and the magnetic field strength in the reconnection inflow region. Assuming a {plasma sheet} of length $L$ with $N$ plasmoids uniformly distributed along it, the ejection timescale for a plasmoid at the edge is $\tau_{eject}\sim L/[(N+1)v_p]$, where $v_p$ is the plasmoid ejection speed. Hence, the magnetic flux transfer rate caused by plasmoid ejection is given by:
	\begin{equation}
	\frac{\phi_{p}}{\tau_{eject}}=\frac{w_pB_0v_p(N+1)}{L}.
\end{equation}	
	
	$\phi_{\eta_m}/\delta t$ is neglectable due to the extremely low efficiency of magnetic flux conversion via Ohmic dissipation (corresponding to a reconnection rate of only $10^{-7}$), and then we have ${\partial \phi_{inflow}}/{\partial t}$=$\phi_{eject}/\delta t\sim \phi_{p}/\tau_{eject}$. Combining Equation (4) with Equation (12) {and neglecting the Ohmic dissipation term ($\eta_m \nabla^2\phi$)}, we obtain: 
		\begin{equation}
		v_{in} \sim	\frac{w_pv_p(N+1)}{L}
	\end{equation}

A newly formed plasmoid undergoes acceleration by magnetic tension in the {plasma sheet} until its speed reaches the Alfvén speed $v_A$, which represents its speed upper limit. Given an acceleration of $a$=5-10 km s$^{-2}$\citep{2008A&A...477..649B} in the corona, the time required to accelerate a plasmoid to the coronal Alfvén speed is only tens of seconds. This means it can reach the speed limit $v_A$ over a very short distance. It then follows that upon ejection from the {plasma sheet}, $v_p \sim v_A$ holds. According to Equation (13), then we obtain: 
		\begin{equation}
		\frac{v_{in}}{v_p} \sim	\frac{v_{in}}{v_A} = M_A \sim \frac{w_p(N+1)}{L},
	\end{equation}
where $M_A$ is the Alfvén Mach number, representing the reconnection rate. Equation (14) indicates that the reconnection rate induced by plasmoid ejection is governed by the plasmoid number $N$, width $w_p$ and the main {plasma sheet} length $L$, thus revealing the role of plasmoid ejection in modulating the reconnection rate. This equation closely matches to the analytical solution for the plasmoid-mediated reconnection rate derived by \citet{2010PhRvL.105w5002U}, and and the two expressions asymptotically converge when $N$ $\gg$ 1 ($N+1\sim N$). Notably, the key parameters—the plasmoid number $N$, width $w_p$ and the main {plasma sheet} length $L$—are all observable. We will subsequently employ Equation (14) with observational measurements to quantify the reconnection rate induced by plasmoid ejections.

 Figure 4c, d and e, f shows the evolution of two observational cases and their temperature before and after the plasmoid ejection, respectively, with key parameters ($\frac{w_p}{L}$, $N$) given in the top-left labeled of the former panels. Substituting these parameters into Equation (14) yields reconnection rate $M_A$= 0.5 and 0.44 for each case, respectively. The order of calculated $M_A$ indicates that plasmoid ejections substantially increase the reconnection rate, driving it into a fast regime. This enables rapid magnetic energy conversion in the reconnection region, which explains the rapid rise of the heated area observed in the {plasma sheet} following plasmoid ejections. This finding is consistent with numerical simulations, theory, and observations of plasmoid-mediated reconnection \citep{2022NatCo..13..640Y,2009PhPl...16k2102B,2010PhRvL.105w5002U,2017ApJ...849...75H}.  Equation (14) essentially describes the reconnection rate of secondary {plasma sheets} modulated by plasmoid ejections. How can we understand this modulation? In the process of a solar eruption, the erupting magnetic flux rope (corresponding to a zero-order plasmoid ejection) intensifies reconnection rate through stretching the main {plasma sheet}, thereby increasing the inflow speeds. Similar to the zero-order plasmoid ejection, the ejection of first-order and higher-order plasmoids can also {enhance} magnetic reconnection rate through stretching their respective secondary {plasma sheets}. This process is self-similar to the zero-order plasmoid ejection, as described by \citet{2001EP&S...53..473S} and \citet{2010PhRvL.105w5002U}. We must note, however, that considering the approximations and the differences between 2D and 3D reconnection, the reconnection rate calculated here should be treated as a reference value within this specific context. It helps demonstrate the role that secondary plasmoid ejections play in increasing the reconnection rate in a torn {plasma sheet}.

\subsubsection{Energy flux induced by plasmoid ejection }

An underlying dependence emerges from the comparison of Figures 4e with 4f: the peak temperature of the heated {plasma sheet} is higher when the ejected plasmoid is larger. In this section, we will derive the Poynting flux during the plasmoid ejection to demonstrate this dependence. The combination of Equations (9) with (13) yields the Poynting flux $S$: 
\begin{equation}
	S \propto \frac{B^2_0 w_pv_p(N+1)}{4\pi L}.
\end{equation}
 This equation provides description for the Poynting flux induced by the plasmoid ejection. Assuming a {plasma sheet} of length $L$ containing $N$ uniformly distributed plasmoids of comparable size, the plasmoid ejection timescale is given by $\tau_{eject}$$= L/[(N+1)v_p]$. Substituting $\tau_{eject}$ into Equation (15), the energy released per plasmoid ejection $E$ can be expressed as:
 \begin{equation}
 	E =S\tau_{eject} \propto \frac{w_p B^2_0}{4\pi},
 \end{equation}
 where $w_p$ and $B_0$ are the plasmoid width and the magnetic field strength of the reconnection inflow region. A dependence indicated by this equation is that the released energy per plasmoid ejection monotonically increases with the plasmoid width, thus providing a physical explanation for the higher peak {plasma sheet} temperature associated with larger ejected plasmoids (see Figures 4e and 4f). This dependence is very intuitive, since the ejection of a larger plasmoid represents the conversion of a greater amount of magnetic free energy, as is often seen in coronal mass ejections \citep{1994Natur.371..495M,1995ApJ...451L..83S}. {It should be noted that Equation (16) only represents a fraction of the total magnetic energy conversion. The magnetic reconnection process includes various energy components \citep{2019ApJ...883..124Z}, such as particle accelerations \citep{2024ApJ...971...85C,2022Natur.606..674F,2006Natur.443..553D} and waves} \citep{2012ApJ...753...52L,2025ApJ...994....6H,2016ApJ...823..150T,2025ApJ...982..142R}.

\section{Drivers for {Plasma Sheet} Formation and Heating}

In this section, we use the aforementioned observational evidence to identify the drivers behind the formation and heating of the {plasma sheet} and to summarize its key observational characteristics.

\textbf{Role of magnetic flux emerging}. Throughout the development of the {plasma sheet}, the key observational characteristics from the photosphere to the corona sequentially include: the continuous emergence of photospheric magnetic flux, the formation and elongation of a curved {plasma sheet}, the rapid uplift of the {plasma sheet}, the repeated appearance of magnetic plasmoids, and jets ejected into the high corona. To analyze the physical relationships among these features, composite images are created (Figure 5) by superimposing SDO/AIA 171 \AA\ maps and synchronal temperature maps onto the corresponding SDO/HMI maps. It should be noted that representative magnetic field lines are artificially added to the figure to aid in understanding the magnetic connectivity between the photospheric magnetic flux and the coronal {plasma sheet}.

The emerging reconnection model provides a coherent framework for understanding the relationships between these observational characteristics, as follows: (1) according to the emerging reconnection model, when emerging magnetic fluxes encounter an oppositely directed ambient field, a reconnecting {plasma sheet} (hereafter the main {plasma sheet}) forms at the X point (hereafter the main X point) between them (see Figure 5a); (2) As a result of the continual magnetic flux emergence, the increased stress at the main X point not only lengthens but also uplifts the main {plasma sheet} (see Figure 5b), a scenario robustly supported by analytical, numerical, and observational studies \citep{2017ApJ...839...22H,2018ApJ...862L..24P,1995Natur.375...42Y,2017ApJ...841...27N}; (3) The lengthened {plasma sheet} promotes the growth of the tearing-mode instability through increasing tearing wavenumber \citep{2017ApJ...849...75H,2009PhRvL.103j5004S,2022A&A...666A..28S}, thereby tearing the main sheet into a chain of plasmoids separated by secondary X points (see Figure 5b); these X-points, in turn, host the secondary {plasma sheets} where secondary reconnection occurs; (4) The emerging magnetic loops at a greater (lesser) height exhibit lower (higher) magnetic tension. Such a non-uniform tension distribution gives rise to an increasing susceptibility of loop deformation with height, thereby forcing the {plasma sheet} into a curved shape \citep{1976SoPh...48...89T}, as shown in Figure 5b. The photospheric flux emergence acts as a collective facilitator that orchestrates the {plasma sheet}'s formation, its lifting and tearing, as well as the coronal jet.

\textbf{Role of plasmoids in {Plasma Sheet} Heating}. The temperature profile of the reconnection region revealed by Figure 5c shows that the two cusps ($\sim 10^{6.8}$ K) is significantly hotter that the main X point. {The best-fitted solution temperature at the main X-point is $10^{6.26}$ K (uncertainty range: $10^{6.18}$ K -- $10^{6.33}$ K; see Figure 5e)}. This temperature difference signifies that it is primarily in the reconnection outflow region that the heating process occurred, rather than in the {plasma sheet} where reconnection proceeds. The temperature difference stems from a magnetic Y-point existing in the outflow region—a site conducive to heating and acceleration mechanisms such as shocks, betatron processes, and Fermi acceleration \citep{2013ApJ...767..168L,2024ApJ...971...85C,1994Natur.371..495M,1983SoPh...84..169F,2012ApJ...753...28G,2023RAA....23c5006L}.

Figure 5d shows that the main {plasma sheet was heated obviously after undergoing the tearing process, with the peak temperature reaching to $\sim 10^{6.85}$ K (uncertainty range: $10^{6.62}$ K -- $10^{6.96}$ K; see Figure 5f), significantly hotter than its pre-tearing temperature of $\sim 10^{6.26}$ K (see Figure 5e)}. Moreover, Figure 5d reveals the fine structure of the main {plasma sheet}, which has fragmented into multiple plasmoids separated by secondary X-points. A zoom-in view indicates that the plasmoids are hotter than the secondary X-point. This temperature difference identifies plasmoids, rather than the secondary X-points, as the primary sites of heating within the main {plasma sheet}. Physically, these secondary X-points represent the sites of secondary {plasma sheets}, which tear the main {plasma sheet} into plasmoids via localized reconnection. The plasmoids themselves constitute the outflow regions of this secondary reconnection (see \citet{2001EP&S...53..473S} for details) and thus host secondary Y-points. These secondary Y-points are self-similar to the primary Y-point of the main {plasma sheet} (Figure 5c), suggesting that plasmoids and the main cusps share the same underlying heating and acceleration mechanisms mentioned before. This direct mechanistic link therefore accounts for why plasmoids are the primary contributor for the heating of the main {plasma sheet}.

\section{Conclusions and Discussions}

This study investigates the connection between photospheric flux emergence and the development of a coronal {plasma sheet} using high-resolution extreme ultraviolet (EUV) imaging and magnetic field data from the Solar Dynamics Observatory (SDO). Our analysis demonstrates that the photospheric magnetic flux drives the formation, elongation, and tearing of the {plasma sheet}. The results underscore the fundamental role of emerging photospheric flux in the evolution of the {plasma sheet} within the atmosphere. Furthermore, the exceptional quality of the observational data provides deep insight into the dual role of plasmoids: their heating effect within the {plasma sheet} and their speedup for the magnetic reconnection process. The key findings of this work are summarized below.

\textbf{Two heating mechanisms}. In this event, the {plasma sheet} did not maintain a persistently high temperature but was intermittently heated, driven by a continuous counterbalance between heating and radiative cooling. The fact that the {plasma sheet} could be heated to million-degree temperatures within short time intervals indicates the presence of efficient heating processes, and four representative heating cases are summarized Figure 3. These four instances can be categorized into two distinct heating mechanisms. The first involves the coalescence of plasmoids with other magnetic structures, including both plasmoid-plasmoid and plasmoid-cusp mergers, as shown in Figures 3a, b, and c. During this coalescence, the resulting secondary {plasma sheets} play a key role in the heating process \citep{1977PhFl...20...72F,1979PhFl...22.2140P,2011ApJ...733..107K,2012A&A...541A..86K}, responsible for the observed plasmoid-plasmoid (Figure 3a) and plasmoid-cusp (Figure 3b and c) coalescence heating, as well as upward-propagating blobs in the jet (Figure 3c and d). 

Another heating mechanism is the tearing of the main {plasma sheet}. On the one hand, the tearing introduces fast reconnection (a referred $M_A\approx0.12\pm0.02$, see Section 3.4.1), leading to the rapid dissipation of magnetic free energy. On the other hand, secondary Y points formed between secondary {plasma sheets} and resulting plasmoids favors multi heating mechanisms, such as shocks, betatron processes, and Fermi acceleration \citep{2013ApJ...767..168L,2024ApJ...971...85C,1994Natur.371..495M,1983SoPh...84..169F,2012ApJ...753...28G,2023RAA....23c5006L}. For this heating mechanism, the wider and more numerous the plasmoids are, the greater the tearing process released energy, as seen by the comparison of figure 4c and d. Therefore, the width and number of plasmoids serve as key observational indicators for quantifying its released energy, as indicated by Equation (10). {It should be noted, however, that this correlation is derived from only one clear case, and thus its robustness needs to be examined through more observations.}

\textbf{Two fast reconnection mechanisms}. 
Observations show that the main {plasma sheet} can be heated in a short time interval, indicating fast conversion of magnetic free energy. A close inspection of the observations, combined with analytical studies, reveals two fast reconnection mechanisms. The first is the tearing of the main {plasma sheet}. For this mechanism, the reconnection rate scales with the number ($N$) and the widening speed ($v_{p\perp}$) of plasmoids, as described by Equation (8). With $N$=1 and $v_{p\perp}$ = 69 $\pm$ 13 km/s from the case in Figure 3a, we obtain an Alfvén Mach number ($M_A$) of approximately $0.12\pm0.02$, which suggests the occurrence of fast reconnection. This high reconnection rate accounts for the rapid heating that closely follows the observed plasmoid widening. The magnetic flux required for plasmoid widening is supplied by the reconnection inflow, which is converted via secondary {plasma sheets}. Therefore, the calculated $M_A$ value indicates that fast reconnection is occurring in these secondary {plasma sheets}, which are formed by the tearing process. As for this fast reconnection mechanism, the rapid appearance and widening of plasmoids is a key observational indicator for rapid heating of the main plasma sheet.

Another mechanism is ejections of plasmoids, which has been evidenced widely in studies of simulations, theories and observations. \citet{2009PhPl...16k2102B} {suggested} that the plasmoid ejection promote the reconnection rate to 0.1 through simulations with Lundquist number $\sim$ $10^5$. Similar conclusion was validated in observational study of \citet{2022NatCo..13..640Y}. Besides, The analytical study of \citet{2010PhRvL.105w5002U} indicates the rate of plasmoid-mediated reconnection ranges from 0.01-0.1. Similarly, our study provides additional observation evidence for the fast reconnection induced by plasmoid ejections. For this mechanism, our result shows that the reconnection rate scales with the plasmoid number ($N$) and the ratio of the plasmoid width to the main {plasma sheet} length ($w_p/L$), as described by Equation (14). Combining observational measurements with analytical studies, we obtain Alfvén Mach numbers of approximately 0.5 and 0.44 with two case (see Figure 4e and 4f). These two values validate fast reconnection, accounting for the fast heating of the main {plasma sheet} closely following plasmoid ejections. An underlying scenario is that ejections of resulting plasmoids stretch secondary {plasma sheets} and thereby facilitate reconnection rate in a way similar to the stretching of the main {plasma sheet} by eruptive magnetic flux ropes (0 order plasmoid; see \citet{2001EP&S...53..473S}). This means the self-similarity between 0 and higher order plasmoid ejections.

A key conclusion available for both two aforementioned fast reconnection mechanisms is: larger plasmoids are associated with a higher reconnection rate and a greater release of energy. This indicates that plasmoid width is a crucial observational indicator for the magnetic reconnection process. Furthermore, owing to the self-similar nature of plasmoid cascading in magnetic reconnection, this conclusion applies not only to secondary plasmoids but also to magnetic flux ropes (considered 0th-order plasmoids) erupting.

As suggested in Figure 2, the development of the main {plasma sheet} appears as two distinct phases separated by a transition at 20:11 UT, characterized by four consistent shifts: a slowed ascent (Figure 2b), a switch to shortening in length (Figure 2c), a significant temperature increase (Figure 2d), and more frequent plasmoid ejections (Figure 2e). The four consistent shifts leads to a key question: Do they share a common underlying driver or just occur casually? We will qualitatively address this below. After 20:11 UT, Figure 2e shows a significant increase in the frequency of plasmoid ejections (indicated by the stripe frequency) and a larger ejection scale (reflected by the stripe width). According to Equation (16), larger-scale ejections release more energy, which explains the higher temperature of the {plasma sheet} after 20:11 UT (Figure 2d). Besides, since the magnetic flux carried by a plasmoid is given by $\phi_p\sim w_p B_0$, the higher frequency and larger scale of plasmoid ejections after 20:11 UT imply a higher magnetic flux conversion rate than before. This suggests that the magnetic flux accumulated in the inflow region by photospheric emergence rapidly diminished after 20:11 UT. Consequently, the magnetic stress in the inflow region weakened, leading to a slower uplift and shortening of the {plasma sheet}. Therefore, the post-20:11 UT shortening of the {plasma sheet}, the reduction in its ascent speed, and its temperature increase are not coincidental but are all driven by the more frequent and larger-scale plasmoid ejections. However, a key question remains yet: why did the scale and frequency of plasmoid ejections increase so markedly after 20:11 UT? We hypothesize that a potential mechanism is the ejection of a large-scale plasmoid around the phase transition. \citet{2010PhRvL.105w5002U} predicted that the ejection of ``monster plasmoids”--those with widths approaching a significant fraction (e.g., ~0.05) of the current sheet length--can lead to abrupt events.  Consistent with this theory, our observations reveal a plasmoid of substantial size (width $\approx$ 0.25L, where L is the current sheet length; see Figure 4c) ejected near the transition, which is seen in appearance of a relatively-wide stripe in Figure 2e {and its zoom-in}. Such consistency with the prediction by \citet{2010PhRvL.105w5002U} provides a possible link bridging the phase transition of the tearing frequency. {However, it should be noted that this link remains hypothetical, as solid observational evidence is still lacking.}

It is crucial to clarify that the observed phase transition reflects a change in the dynamics of the plasma sheet (e.g., the length, height, temperature and tearing frequency), not necessarily the exact onset of fast magnetic reconnection. While the precise timing of fast reconnection onset is difficult to pinpoint observationally in this event, several indicators suggest it was already ongoing prior to the phase transition. For instance, at 19:38 UT, the plasma is heating rapidly (see animated Figure 1), coinciding with plasmoid formation. Although lacking quantitative analysis, these indicator strongly suggest occurrence of fast reconnection. 

\begin{acknowledgments}
 The authors wish to thank the anonymous referee for insightful suggestions, which have significantly strengthened this work. We gratefully acknowledge helpful discussions with Pro. Lei Ni and Xiaoli Yan from Yunnan Observatories, and also the science teams of the Solar Dynamics Observatory (SDO) for their open data policy and the high-quality observations used herein. This work is sponsored by the National Key R\&D Program of China under No. 2024YFA16112001, the Strategic Priority Research Program of the Chinese Academy of Sciences under No. XDB0560000, the National Natural Science Foundation of China (NSFC)  under  No. 12325303, Yunnan Key Laboratory of Solar Physics and Space Science under the number No. 202205AG070009, and Yunnan Fundamental Research Projects under the numbers Nos. 202301AT070347 and 202301AT070349.
\end{acknowledgments}

 \appendix
 \section{DERIVATIONS FOR INFLOW FLUX}
The injection rate of magnetic fluxes in the reconnection region is described by 
  \begin{equation}
 	\frac{\partial \phi}{\partial t} = -\vec{v} \cdot \nabla \phi+\eta_m \nabla^2\phi,
 \end{equation}
 where $\phi$ and $\eta_m$ denote the magnetic flux and magnetic diffusivity, respectively. The magnetic diffusivity $\eta_m$ depends on the plasma resistivity. We thus obtain
 
   \begin{equation}
	\frac{\partial \phi}{\partial t} = -({v_x}\frac{\partial \phi}{\partial x}+v_y\frac{\partial \phi}{\partial y}+{v_z}\frac{\partial \phi}{\partial z})  +  \eta_m \nabla^2\phi.
\end{equation}
 The magnetic field strength is given by $\vec{B}=\nabla \times (\phi \hat{z})$, which yields
 
\begin{equation}
	\vec{B}=(\frac{\partial \phi}{\partial y},-\frac{\partial \phi}{\partial x},0).
\end{equation}
Consider reconnection confined to the two-dimensional (x–y) plane, with the magnetic field directed along the y‑axis. Then combining Equations (A2) and (A3) yields
    \begin{equation}
 	\frac{\partial \phi}{\partial t} = {v_x}B_y  +  \eta_m \nabla^2\phi.
 \end{equation}
 Equation (A4) can be alternatively expressed as 
 \begin{equation}
\frac{\partial \phi}{\partial t} = {v_{in}}B_0  +  \eta_m \nabla^2\phi
 \end{equation}
 where $ {v_{in}}$ and $B_0$ respectively are the inflow speed and magnetic field strength in the inflow region. In plasmoid-mediated reconnection that develops from a Sweet-Parker reconnection, the term  ${v_{in}}B_0$ and $\eta_m \nabla^2\phi$ can be attributed to the injection rate of magnetic fluxes induced by the tearing instability \citep{2010PhRvL.105w5002U} and magnetic diffusivity \citep{1957JGR....62..509P}, respectively.

\bibliographystyle{aasjournal}
\bibliography{bibfile}

\begin{figure}
\epsscale{1}
\figurenum{1}
\plotone{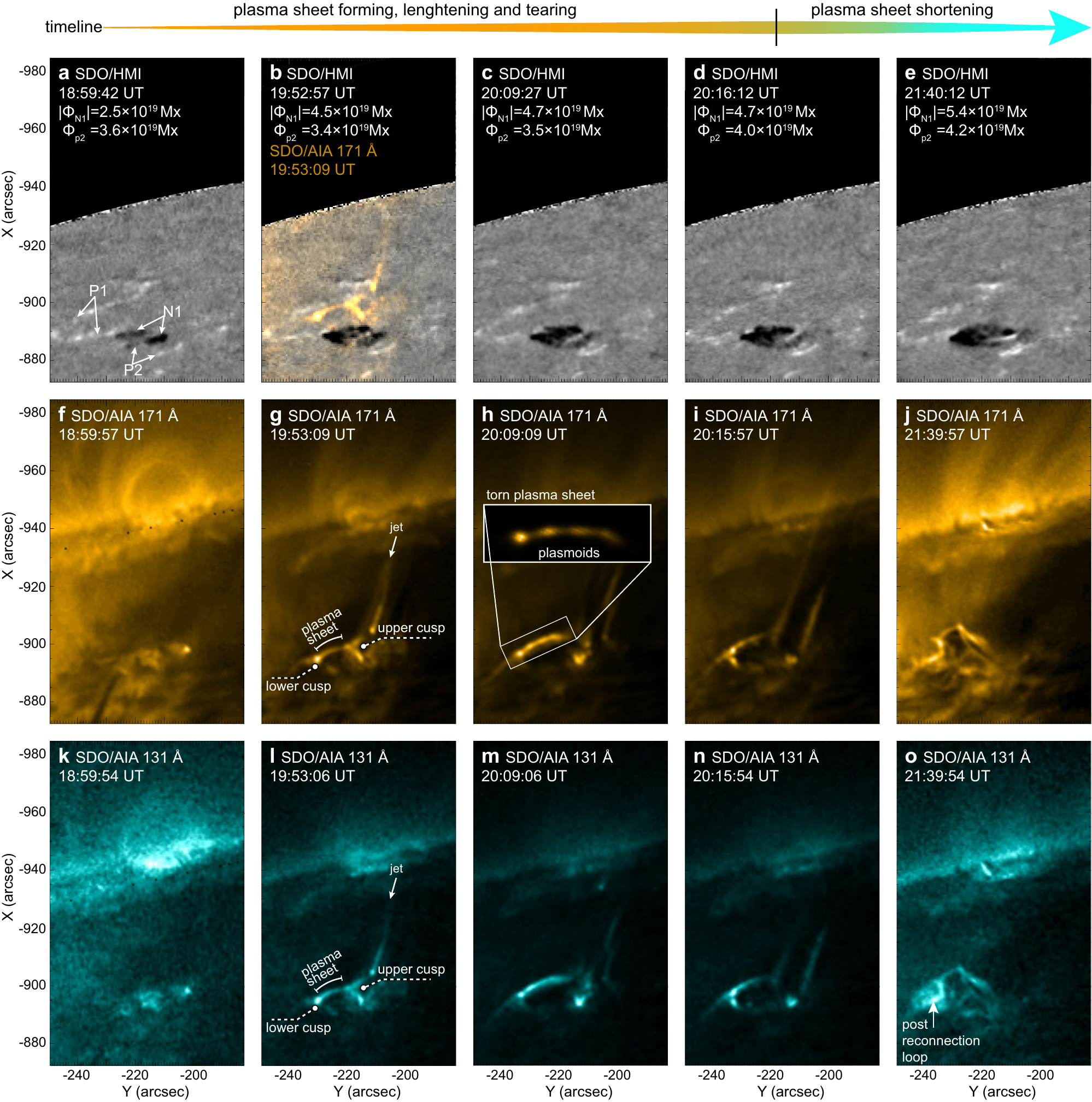}
\caption{The layered atmospheric evolution during magnetic reconnection. From top to bottom are observations from SDO/HMI, SDO/AIA 171 \AA, and SDO/AIA 131 \AA. Time runs from left to right. The temporal sequence illustrates the photospheric magnetic flux emergence process (first row) and the synchronous evolution of coronal atmospheric reconnection activities (second and third rows), including {plasma sheet} formation, lengthening, tearing, and shortening. Representative structures or information are labeled in the corresponding panels, including key magnetic polarities (a; P1, P2, and N1), instantaneous magnetic flux (a-e), coronal jets (g and l), {plasma sheets} (g and l), lower and upper cusps (g and l), plasmoids (h), torn {plasma sheets}, and post-reconnection loops (o). In panel b, the SDO/AIA 171 Å image is overlaid on the SDO/HMI image from panel g to examine the connectivity between the atmospheric reconnection field lines and the photospheric polarities. An animation spanning 19:00--22:00 UT is available online, including the observations of the SDO/HMI, averaged temperature, SDO/AIA 171 \AA\ and SDO/AIA 131 \AA.
\label{fig1}}
\end{figure}

\begin{figure}
\epsscale{1.1}
\figurenum{2}
\plotone{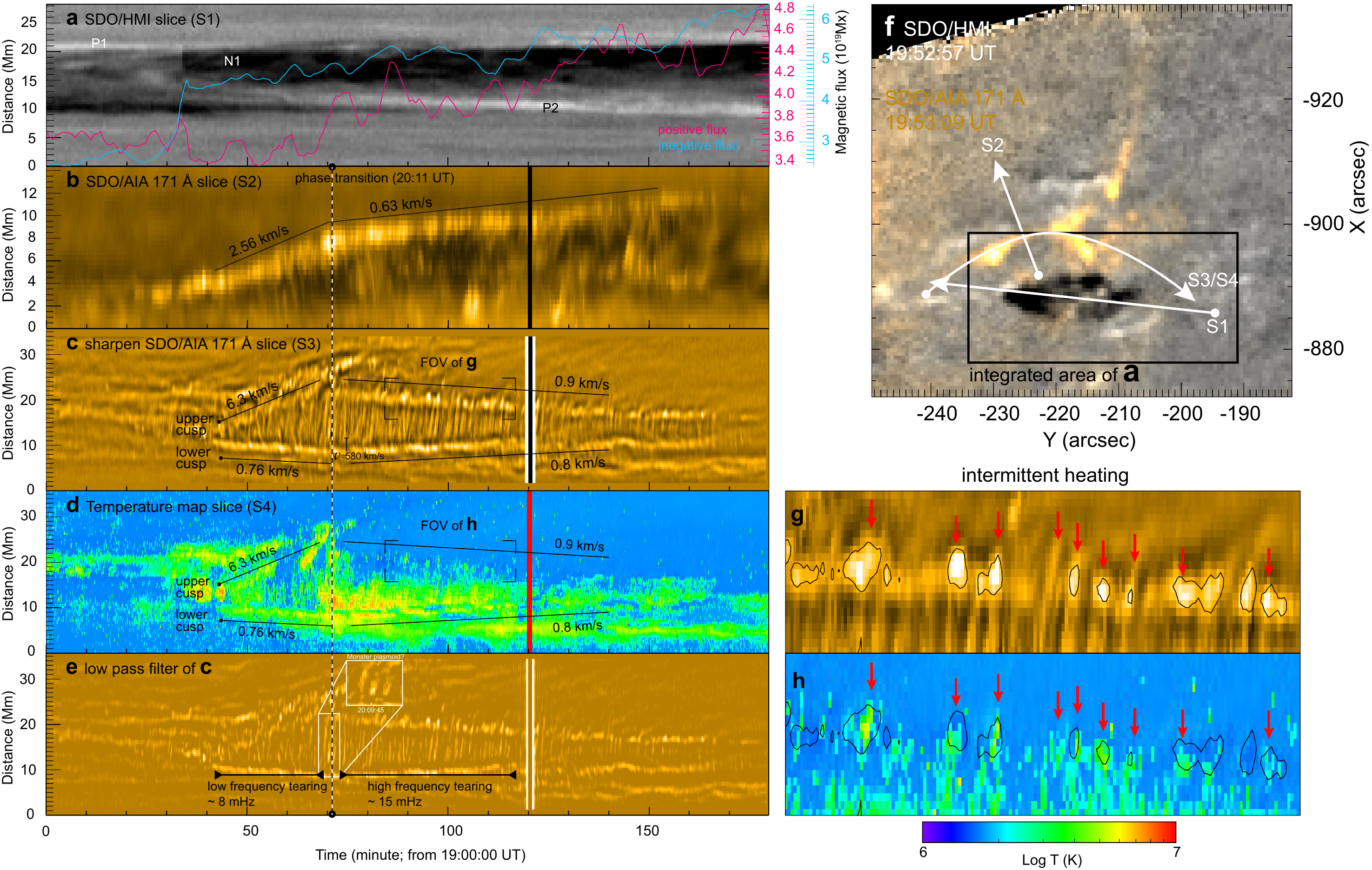}
\caption{The photospheric magnetic flux evolution and the development of the coronal {plasma sheet}. Panels (a) to (d) are slices respectively showing temporal-spatial trajectories of photospheric footpoints, {plasma sheet} height, {plasma sheet} length, and {plasma sheet} temperature, with their path shown in (f). Note that (f) only show the path of S3 and S4 in a fixed frame, which indeed ascends in step with the rising of the main {plasma sheet} in animated frames. Panel (a) includes the integrated photospheric magnetic flux curves of the source region, with the integration area shown in (f); {all slices have a sampling width of 7 pixels}. Panel (e) shows a low-pass filtered version of panel (c) to further shown tearing progress of the {plasma sheet}. Panels (g) and (h) are zoom-in versions of panels (c) and (d), respectively, with the field of view (FOV) indicated by the dashed boxes in (c) and (d); {the contours from panel (g) are overlaid on panel (h) to facilitate comparison}. Key observed structures are labeled in each panel, including footpoints P1, N1, and P2 in (a), the {plasma sheet} ascent speed in (b), the cusp structure and {plasma sheet} expansion speed in (c) and (d). An animation spanning 19:00--22:00 UT is available online, including the observations of the SDO/AIA 171 \AA\ and its synchronous slice.
\label{fig2}}
\end{figure}

\begin{figure}
\epsscale{1}
\figurenum{3}
\plotone{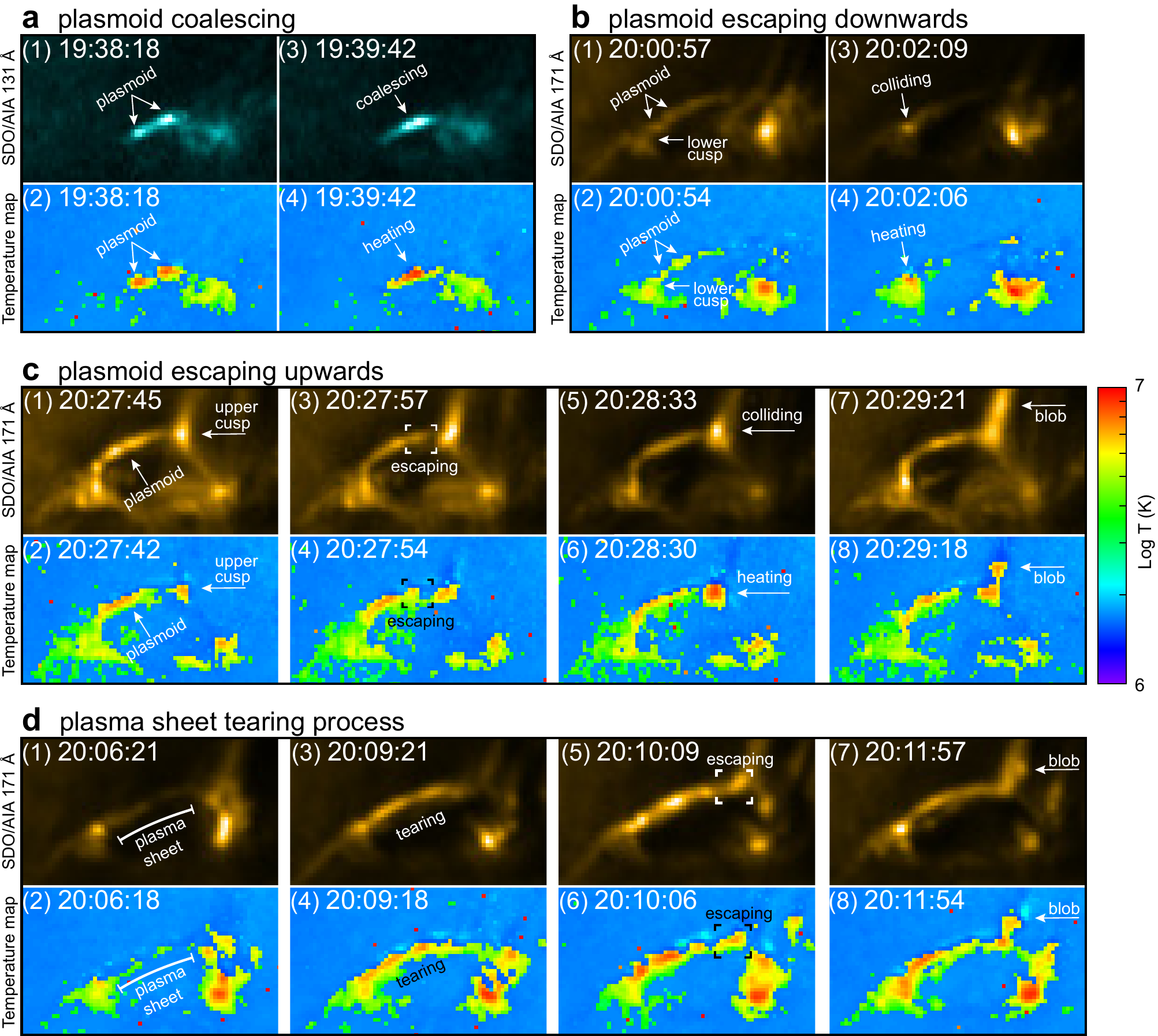}
\caption{Case-by-case analysis of heating progresses. (a) to (d) show four heating cases, namely magnetic plasmoid-plasmoid merging process, plasmoid downward escape process, plasmoid upward escape process, and {plasma sheet} tearing process. Each panel consists SDO/AIA extreme ultraviolet (EUV) evolutionary images and synchronal temperature maps. Key structures are annotated in each panel.
\label{fig3}}
\end{figure}

\begin{figure}
	\epsscale{1}
	\figurenum{4}
	\plotone{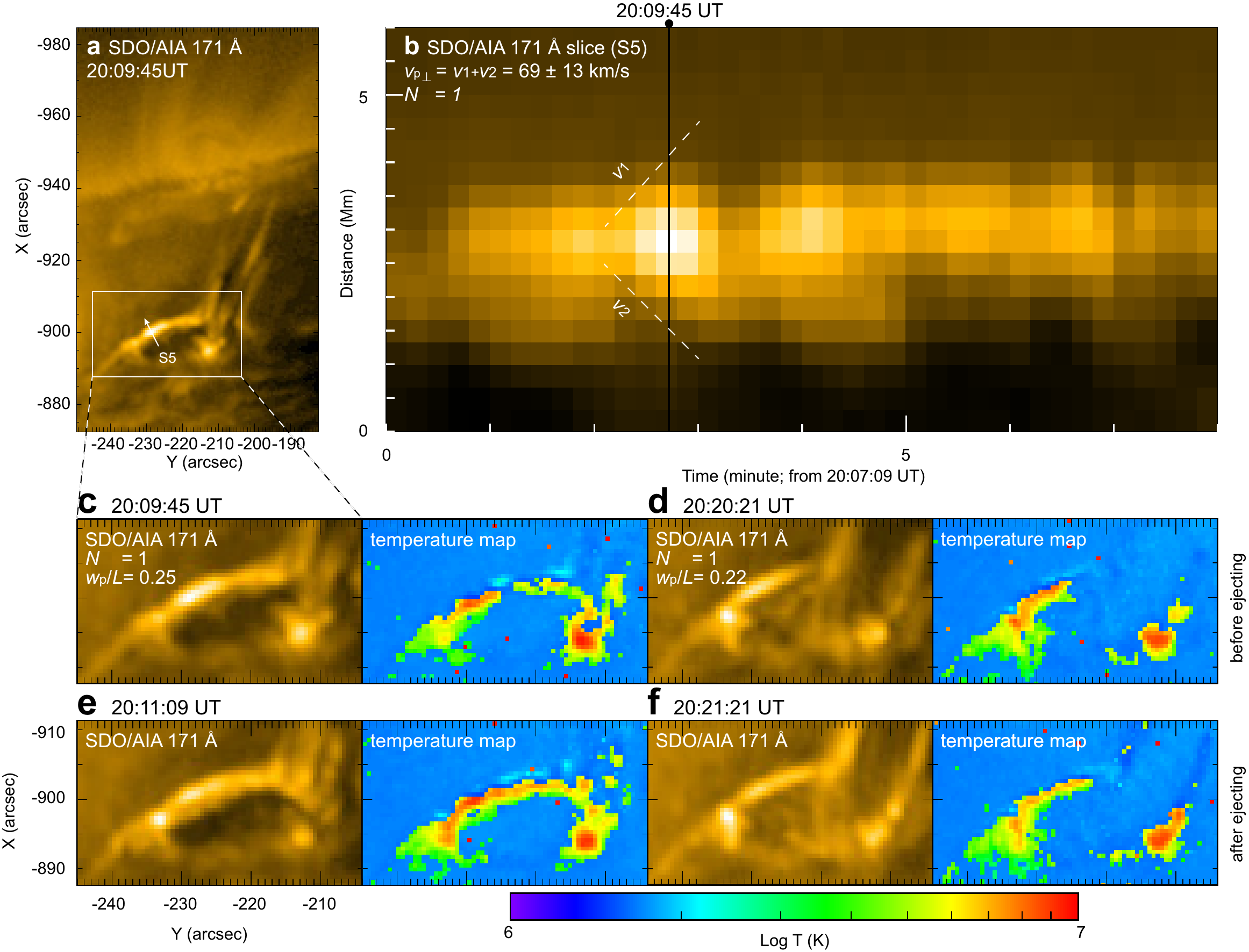}
	\caption{The details of how the formation and ejection of plasmoids on different scales modulate the {plasma sheet} heating. (a) shows a fixed frame of the plasmoid formation at 20:09:09 UT. (b) presents a time slice of the plasmoid widening process, with its path indicated in panel (a); {the slice has a sampling width of 7 pixels}. (c, d) and (e, f) illustrate the {plasma sheet} heating before the formation (pre-ejection) and after the ejection of plasmoids at different scales. The left and right sides of each panel correspond to SDO/AIA 171 \AA\ observations and temperature maps, respectively. {An animation spanning 20:07--20:15 UT is available online, showing animated versions of panels (a) and (b).}
		 \label{fig4}}
\end{figure}

\begin{figure}
	\epsscale{1}
	\figurenum{5}
	\plotone{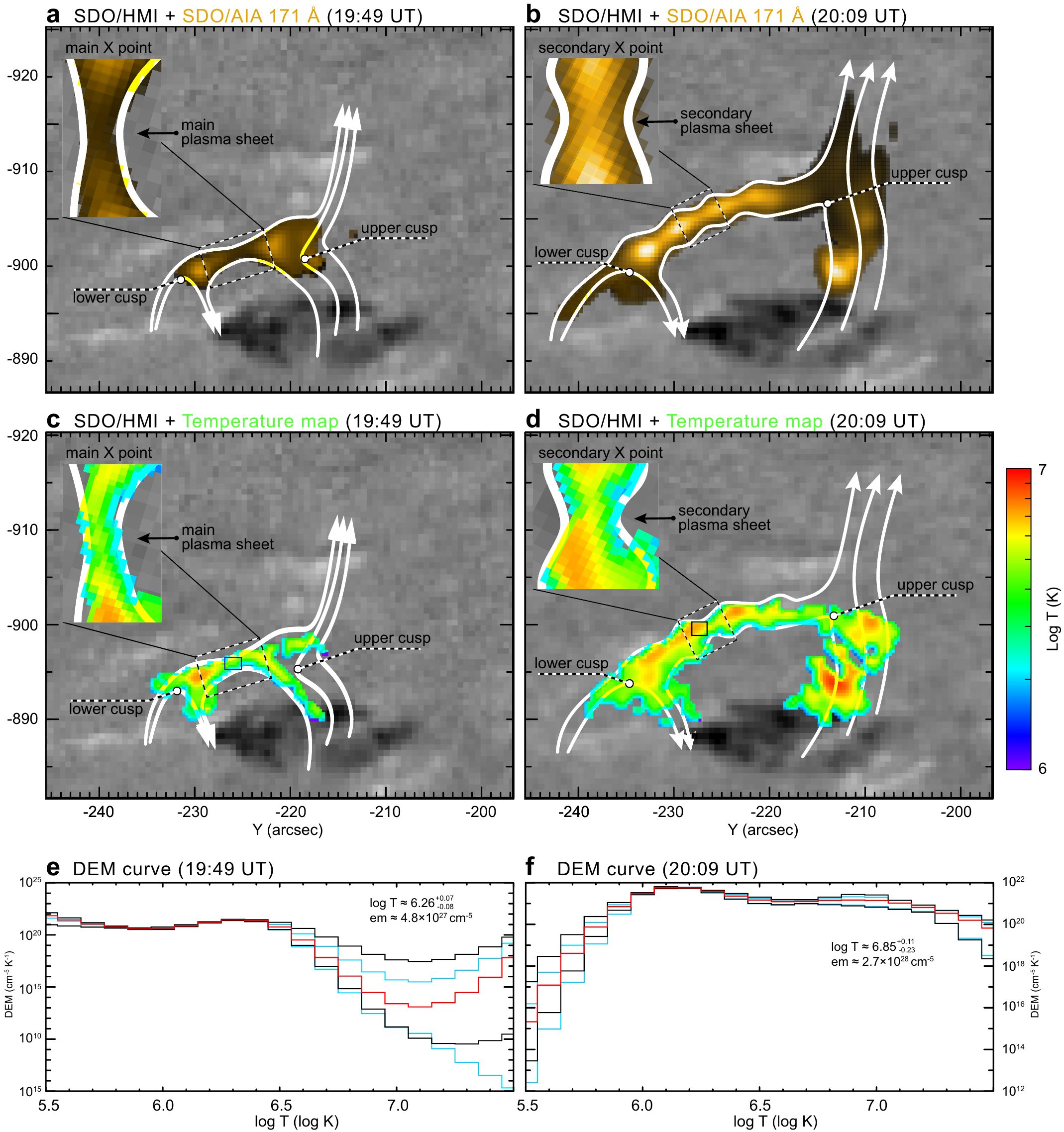}
	\caption{Graphic illustration of the observed event. The first and second rows correspond to composite images of SDO/AIA 171 Å and temperature maps with SDO/HMI images, respectively. From left to right, panels display the observational characteristics before and after the tearing of the coronal {plasma sheet} that was driven by photospheric magnetic flux emerging, including the main {plasma sheet}, secondary {plasma sheets}, and plasmoids. Representative magnetic field lines have been artificially added to guide eyes. {(e) and (f) are DEM curves of regions indicated by black boxes in (c) and (d), respectively; the red lines stands for the best-fitted DEM curves, while the black and blue curves represent the reconstructed curves from the 50 and 100 Monte Carlo (MC) simulations, respectively.}
		\label{fig5}}
\end{figure}

\end{document}